\documentclass[aps,prd,showpacs,eqsecnum,floatfix,twocolumn]{revtex4-1}
\usepackage{amsmath,amssymb,graphicx,color,ulem}

\begin{document}

\title{High resolution numerical-relativity simulations for the merger of
binary magnetized neutron stars} 

\author{Kenta Kiuchi}
\affiliation{Yukawa Institute for Theoretical Physics, 
Kyoto University, Kyoto, 606-8502, Japan~} 

\author{Koutarou Kyutoku} 
\affiliation{Department of Physics,
  University of Wisconsin-Milwaukee, P.O. Box 413, Milwaukee,
  Wisconsin 53201, USA}

\author{Yuichiro Sekiguchi}
\affiliation{Yukawa Institute for Theoretical Physics, 
Kyoto University, Kyoto, 606-8502, Japan~} 

\author{Masaru Shibata}
\affiliation{Yukawa Institute for Theoretical Physics, 
Kyoto University, Kyoto, 606-8502, Japan~} 

\author{Tomohide Wada}
\affiliation{National Astronomical Observatory of Japan, 
Mitaka, 181-8588, Japan~}
\date{\today}

\begin{abstract}

We perform high-resolution magnetohydrodynamics simulations of binary
neutron star mergers in numerical relativity on the Japanese
supercomputer K. The neutron stars and merger remnants are covered by
a grid spacing of 70\,m, which yields
the highest-resolution results among those derived so far.  By an
in-depth resolution study, we clarify several amplification mechanisms
of magnetic fields during the binary neutron star merger for the first time. First, the
Kelvin-Helmholtz instability developed in the shear layer at the onset
of the merger significantly amplifies the magnetic fields. A
hypermassive neutron star (HMNS) formed after the merger is then
subject to the nonaxisymmetric magnetorotational instability, 
which 
amplifies the magnetic field in the HMNS. These
two amplification mechanisms cannot be found with
insufficient-resolution runs. We also show that the HMNS eventually
collapses to a black hole surrounded by an accretion torus which is
strongly magnetized at birth. 

\end{abstract}


\maketitle


\emph{Introduction.}--- Coalescence of binary neutron stars (BNS) is one
of the most promising sources of gravitational waves.  The
second-generation gravitational-wave detectors like advanced LIGO, advanced VIRGO, and
KAGRA~\cite{Detectors}, which will operate in a few years, 
may detect gravitational waves from BNS mergers as
frequently as $\sim 1$--$100$/yr~\cite{Kalogera,RateLIGO}.  If
gravitational waves from BNS mergers are observed, they could tell us
the validity of general relativity in strong gravitational-field regions and
the equation of state (EOS) of neutron stars. Furthermore, a
long-standing puzzle on the central engine of short-hard gamma-ray
bursts (SGRB) may be resolved if gravitational waves are observed
simultaneously with them.  BNS also attracts attention as 
a possible site of r-process nucleosynthesis~\cite{Lattimer} and as 
a source of electromagnetic transients.  In particular, 
emission associated with the radioactive decay of the
r-process elements in the merger ejecta is a promising electromagnetic
counterpart of BNS mergers~\cite{Li:1998bw}. In fact, ``kilonova''
associated with GRB130603B is an interesting candidate of such
events~\cite{GRB130603B}. All these facts stimulate us to theoretically
construct a reliable model of the BNS merger. Numerical
relativity is the unique approach for this purpose.

Strong magnetic fields are universal elements of neutron
stars, shown by pulsar observations~\cite{Manchester}. Typical strength of the magnetic fields is 
$10^{11}$--$10^{13}$\,G.  The so-called magnetars have
even stronger magnetic fields of $10^{14}$--$10^{15}$\,G. Although 
magnetic fields could be a key ingredient in the BNS mergers, their role is still not clear. 
The prime reason is that a number of magnetohydrodynamical instabilities, which can amplify the magnetic
fields, are generally activated by
short-wavelength modes, i.e., the fastest growing mode has a short 
wavelength and is not easily resolved
in numerical simulations.  One example is the Kelvin-Helmholtz (KH)
instability. In the absence of gravity, this 
instability sets in for all the wavelengths, and moreover, the
shorter-wavelength modes have the larger growth rates.  Another example
is the magnetorotational instability (MRI)~\cite{Balbus:1991ay}, in
which the wavelength of the fastest growing mode is quite short for the
typical magnetic-field strength and density of neutron stars.  It
has not been easy to prepare a sufficient grid resolution for the BNS
merger simulations for them~\cite{GRMHD}.

We tackle this problem using the 10 PFLOPS Japanese supercomputer K,
which enables us to assign the highest grid resolution 
so far in this field.  To assess the resolution dependence of the
magnetic-field amplification processes, we carry out an in-depth
resolution study. Furthermore, to 
explore the final state of the BNS merger, we perform longterm
simulations of duration $\sim 100$\,ms. 
Together with the recent observations of $\approx 2M_\odot$ 
neutron stars~\cite{Demorest:2010bx}, 
the recent numerical relativity simulations 
have established that, 
in the BNS mergers for the typical total mass $2.6$--$2.8 M_\odot$ 
and for plausible EOS, 
a hypermassive neutron star (HMNS) is transiently formed after the merger 
and subsequently it collapses to a black hole~\cite{Hotokezaka:2013iia}. 
Based on this picture, we focus in particular on 
the following three stages. First is the stage in which two neutron stars come into
contact. This stage is subject to the KH instability, 
which develops in a thin shear layer~\cite{Price:2006fi}. The second is the HMNS 
phase which is subject to the MRI because of a rapid and strong
differential rotation~\cite{Duez-Shibata:2006}. 

The third is the stage
after the HMNS collapses to a black hole (BH) surrounded by an accretion
torus, which could be again subject to the MRI. Throughout the analysis for the three stages, we clarify the
amplification mechanisms of magnetic fields. 


\emph{Method, initial models and grid setup.}--- Einstein's
equation is solved in the puncture-BSSN formalism~\cite{BSSN}. 
The MHD equation is solved by a high-resolution
shock-capturing scheme with the third-order
cell-reconstruction (see Ref.~\cite{KKS2012} for
details). 
A fixed mesh-refinement algorithm is employed to resolve 
the wide dynamical range of BNS mergers 
simultaneously, where we prepare 7
refinement levels with the varying grid spacing as $\Delta
x_l=2^{7-l}\Delta x_7$ ($l=1, 2, \cdots, 7$) for the same coordinate
origin. Here, $\Delta x_l$ is the grid spacing for the $l$-th level in
the Cartesian coordinates.  The solenoidal constraint and magnetic flux
conservation on the refinement boundary are satisfied using
the Balsara's method~\cite{Balsara,KKS2012}.  
The orbital plane symmetry is imposed. For each level,
the computational domain covers $[-N \Delta x_{l},
N\Delta x_{l}]$ for $x$- and $y$-directions, and $[0, N\Delta x_{l}]$
for $z$-direction.  The highest-resolution runs were done 
with $16,384$~CPUs on the K. 

Table~\ref{tab1} lists the key parameters of our models and numerical
setup of the simulations. We employ H4
EOS~\cite{H4}, with which the maximum mass of neutron stars is
$2.03M_\odot$, and the mass of each neutron star is chosen to be
$1.4M_\odot$. With the parameters chosen, the computation follows about
6 inspiral orbits, and the merger outcome is a HMNS with its lifetime $\sim
10$\,ms in the absence of magnetic fields~\cite{Hotokezaka:2013iia}.  We
prepare three grid resolutions with $\Delta x_7=70$, $110$, and $150$\,m
as well as three maximum initial magnetic-field strengths, $10^{14.5}$,
$10^{15}$, and $10^{16}$\,G to assess how the result depends on the 
resolution and field strength.  The initial magnetic field 
is given in terms of the vector potential
\begin{align*}
A_i = \left( -(y-y_{\rm c})\delta^x_i + (x-x_{\rm c})\delta^y_i\right)
A_{\rm b} [{\rm max}(P-P_{\rm c},0)]^2,
\end{align*}
where $x_{\rm c}$ and $y_{\rm c}$ are the coordinates of the 
stellar centers, $P$ is the
pressure, and $P_{\rm c}$ is the pressure for $\rho=0.04\rho_{\rm max}$.
$A_{\rm b}$ determines the field strength.  The EOS is parametrized by a
piecewise polytrope~\cite{rlof2009} and the $\Gamma$-law EOS is added
during the simulation to take into account the shock heating effect with
the gamma index being 1.8 (see~\cite{Hotokezaka:2013iia} for details).

\begin{table}
\centering
\caption{\label{tab1} Parameters of the BNS and grid setup. 
  $\Delta x_{7}$ is the grid spacing in the finest
  refinement level and $N$ is the grid number in one positive
  Cartesian direction. The last column is the initial maximum strength
  of the magnetic field.  Model name follows the EOS, the initial
  maximum field strength, and grid spacing. 
  The sum of the ADM masses in isolation of each NS $(m_0)$ is 2.8 $M_\odot$ in
  all the models. 
  The initial orbital angular velocity $\Omega$ is set to be 
  $G m_0\Omega/c^3=0.0221$ in all the models with $G$ and $c$ being the
  gravitational constant and the speed of light, respectively.
  }
\begin{tabular}{ccccccccc}
\hline\hline
Model & $\Delta x_{7}$ [m] 
& $N$ & $\rm log_{10}[B_{\rm max}({\rm G})]$\\
\hline
H4B15d70  & 70  & 512  & 15.00 \\
H4B15d110 & 110 & 322  & 15.00 \\
H4B15d150 & 150 & 240  & 15.00 \\
H4B14d70  & 70  & 512  & 14.52 \\
H4B16d70  & 70  & 512  & 16.00 \\
H4B16d110 & 110 & 322  & 16.00 \\
H4B16d150 & 150 & 240  & 16.00 \\
\hline\hline
\end{tabular}
\end{table}

\begin{figure*}[t]
\hspace{-5mm}
\begin{minipage}{0.27\hsize}
\begin{center}
\includegraphics[width=6.0cm,angle=0]{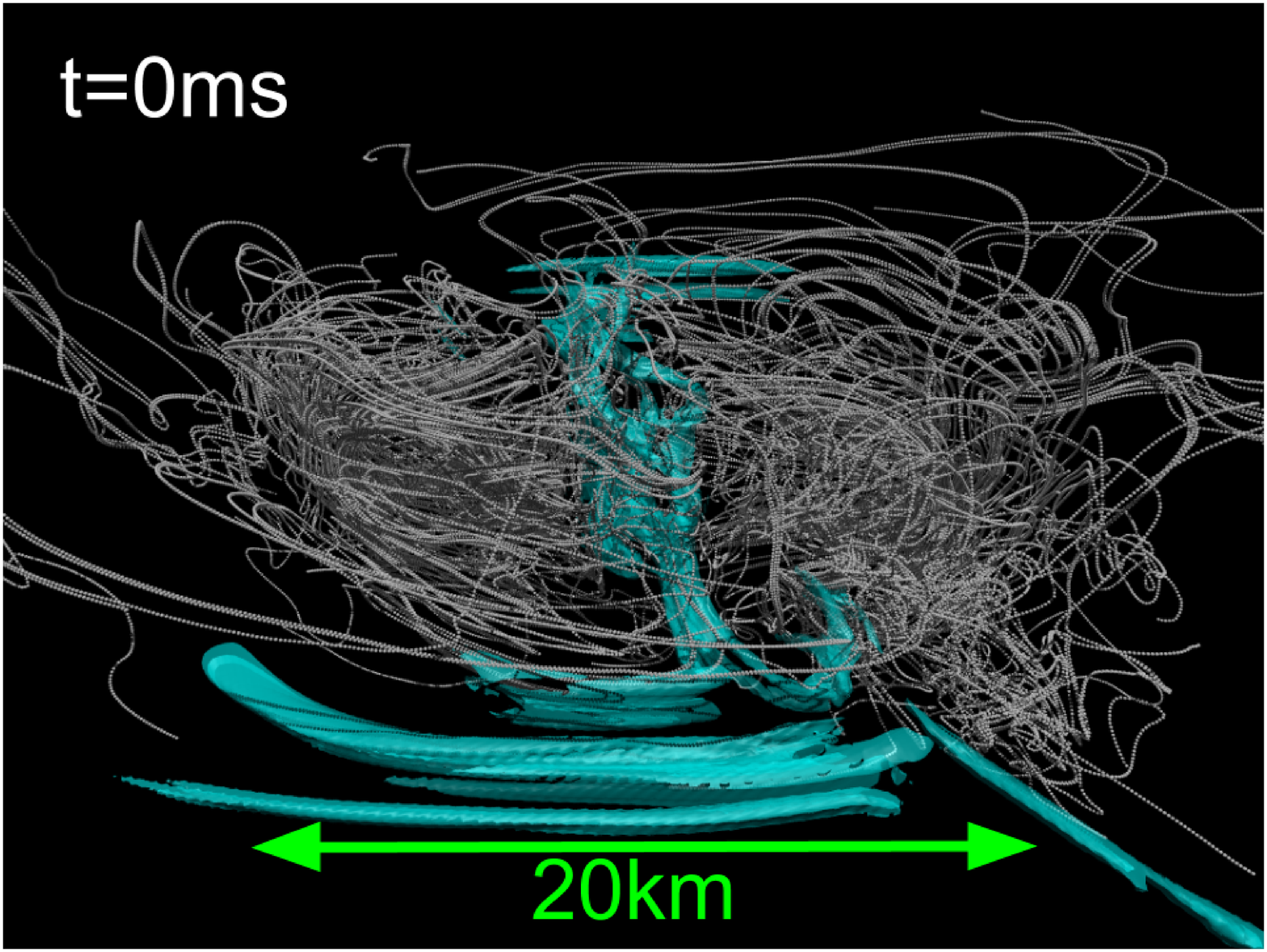}
\end{center}
\end{minipage}
\hspace{10mm}
\begin{minipage}{0.27\hsize}
\begin{center}
\includegraphics[width=6.0cm,angle=0]{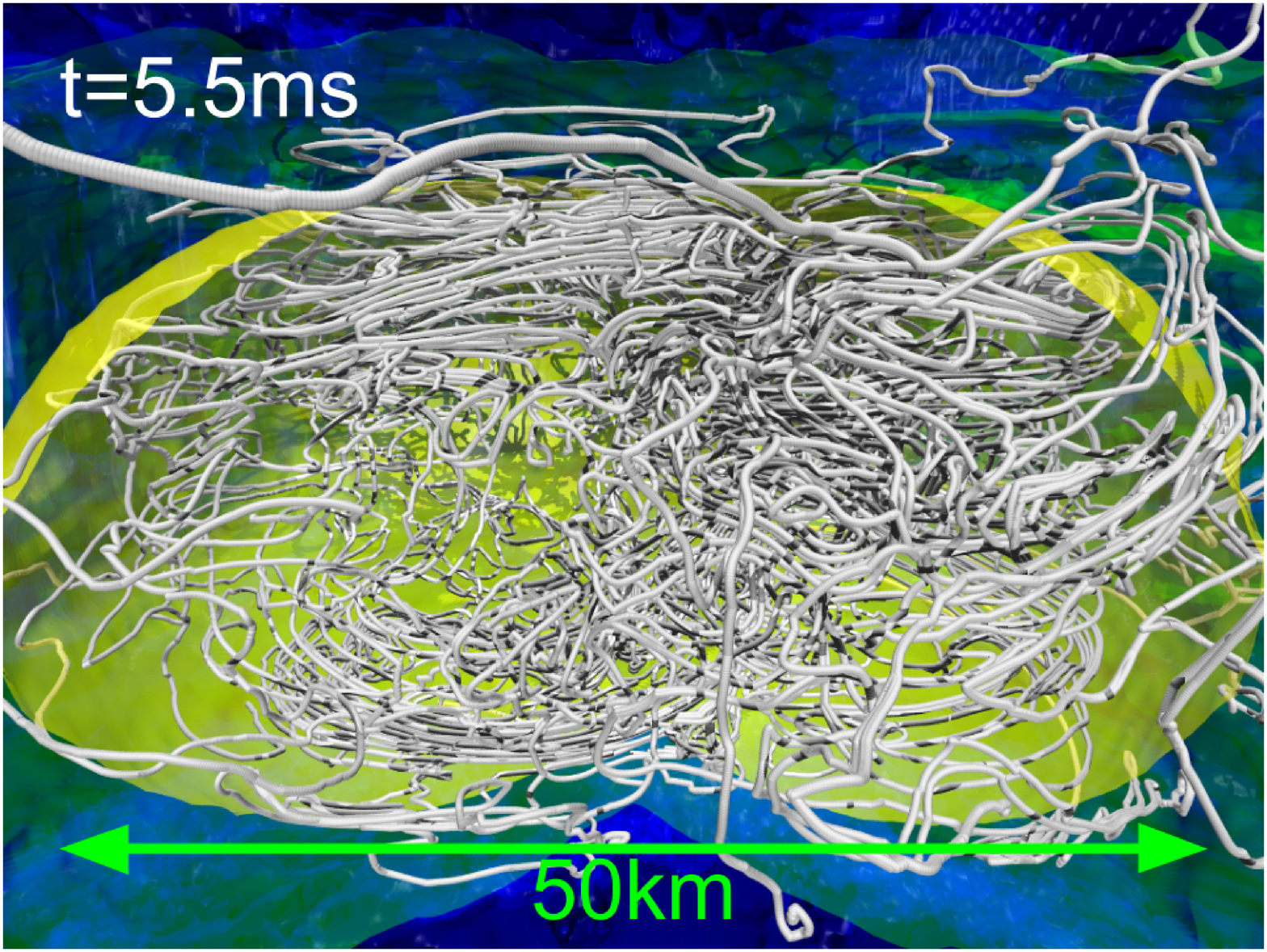}
\end{center}
\end{minipage}
\hspace{10mm}
\begin{minipage}{0.27\hsize}
\begin{center}
\includegraphics[width=6.0cm,angle=0]{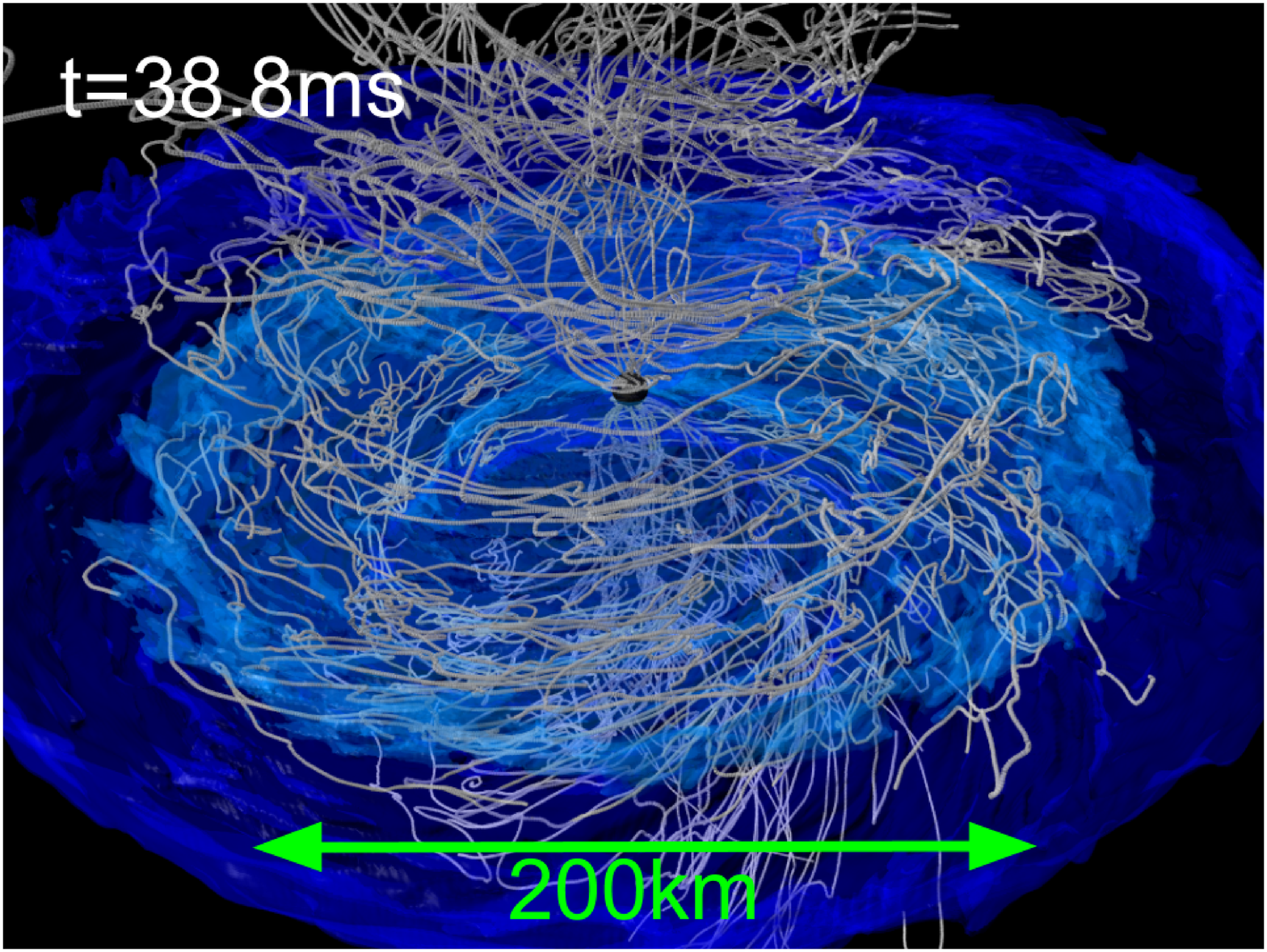}
\end{center}
\end{minipage}
\caption{\label{fig1} Snapshots of the density, magnetic-field 
  strength and magnetic-field lines for H4B15d70 at $t-t_{\rm mrg} \approx 0.0 {\rm
  ms}$ (left panel), at $t-t_{\rm mrg} \approx 5.5 {\rm ms}$ (middle
  panel), and at $t-t_{\rm mrg} \approx 38.8 {\rm ms}$ (right panel). 
  $t_{\rm mrg}$ is a time when the amplitude of the gravitational waves becomes 
  maximum. 
  The left, middle, and right panels show the
  configuration just after the onset of the merger, for the HMNS phase,
  and for a BH surrounded by an accretion torus, respectively.  In each
  panel, the white curves are the magnetic-field 
  lines.  In the left panel, the cyan
  represents the magnetic fields stronger than $10^{15.6}$ G.  In the middle panel,
  the yellow, green, and dark blue represent the density iso-surface of
  $10^{14}$, $10^{12}$, and $10^{10} {\rm g/cm^3}$, respectively.  In
  the right panel, the light and dark blue are the density iso-surface
  of $10^{10.5}$ and $10^{10} {\rm g}/{\rm cm^3}$, respectively.  }
\end{figure*}

\emph{Results.}--- Figure~\ref{fig1} plots the profiles of the density,
magnetic-field strength, and magnetic-field lines at selected time
slices for H4B15d70.  The magnetic fields do not affect the inspiral
dynamics because the magnetic stress-energy is much smaller than the
matter pressure~\cite{GRMHD}.  The left panel shows a snapshot just
after two neutron stars come into contact. In this phase, the KH
vortices develop and curl the magnetic-field 
lines, generating the strong toroidal
fields. This significantly enhances the magnetic stress-energy in the
{\it shear layer}.  The unstable shear layer disappears in a
dynamical time scale of $\sim 0.1$\,ms, because the compression and
resulting shock heating associated with violent oscillations of the
formed HMNS suppress the continuous generation of the vortices.

The middle panel plots a snapshot in the HMNS phase.  This shows that
large-scale toroidal magnetic fields, enhanced primarily by magnetic
winding, are generated.  Furthermore, a detailed analysis elucidates
that the magnetic fields are also globally amplified by the MRI (see below). The
HMNS collapses to a BH at $\approx 14$ ms after the merger 
and a part of the HMNS forms an accretion torus
surrounding the BH.  The non-dimensional BH spin is $\approx 0.69$ and
the torus mass is $\approx 0.06 M_\odot$ at 10 ms after the BH formation
for H4B15d70. These numbers depend slightly on the grid resolution.

The MRI preserves the turbulent flow and
vortices inside the accretion torus and they enhance the 
accretion due to the outward angular-momentum transport. The density of
the accretion torus gradually decreases and $10^{10}$--$10^{11}~{\rm
g/cm^3}$ for $\sim 10$--$30$ ms after the BH formation. The magnetic
field still remains to be toroidal-field dominant, and we do not find
any coherent poloidal field at this moment as shown in the right panel
of Fig.~\ref{fig1}.  This is in contrast to the result of
Ref.~\cite{Rezzolla:2011da}, which reported the formation of a coherent
poloidal field within a relatively short timescale, i.e., $\approx$\,12
ms after the BH formation.  It is not trivial to generate such
a coherent poloidal field. A large amount of matter
is ejected and blown outwards in the merger phase and the resulting ram 
pressure due to the fall-back toward the BH and torus suppresses the matter outflow. 
Since the magnetic-field lines are frozen in the fluid
elements, an outflow 
which has not been seen for $t-t_{\rm mrg}\lesssim 40$ ms will be 
necessary to generate a coherent poloidal magnetic field. 

\begin{figure}[t]
\begin{center}
\includegraphics[width=7.0cm,angle=0]{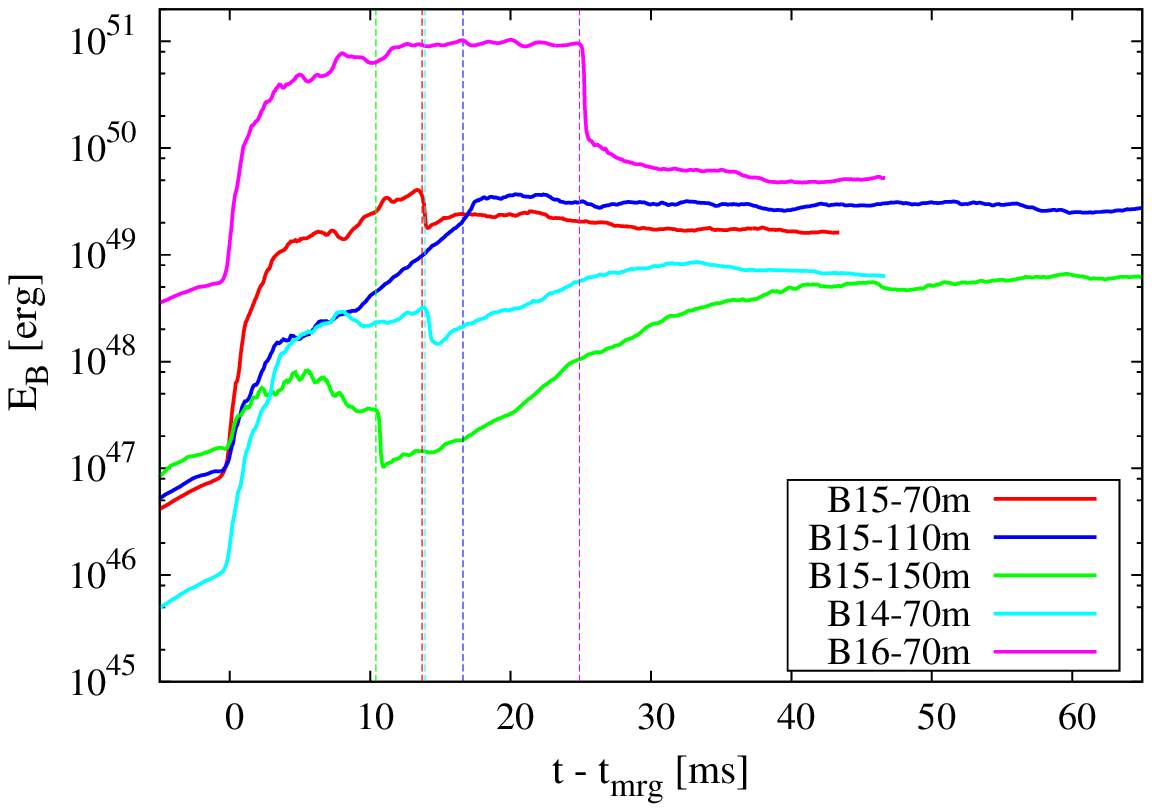} \\
\vspace{-10mm}
\includegraphics[width=7.0cm,angle=0]{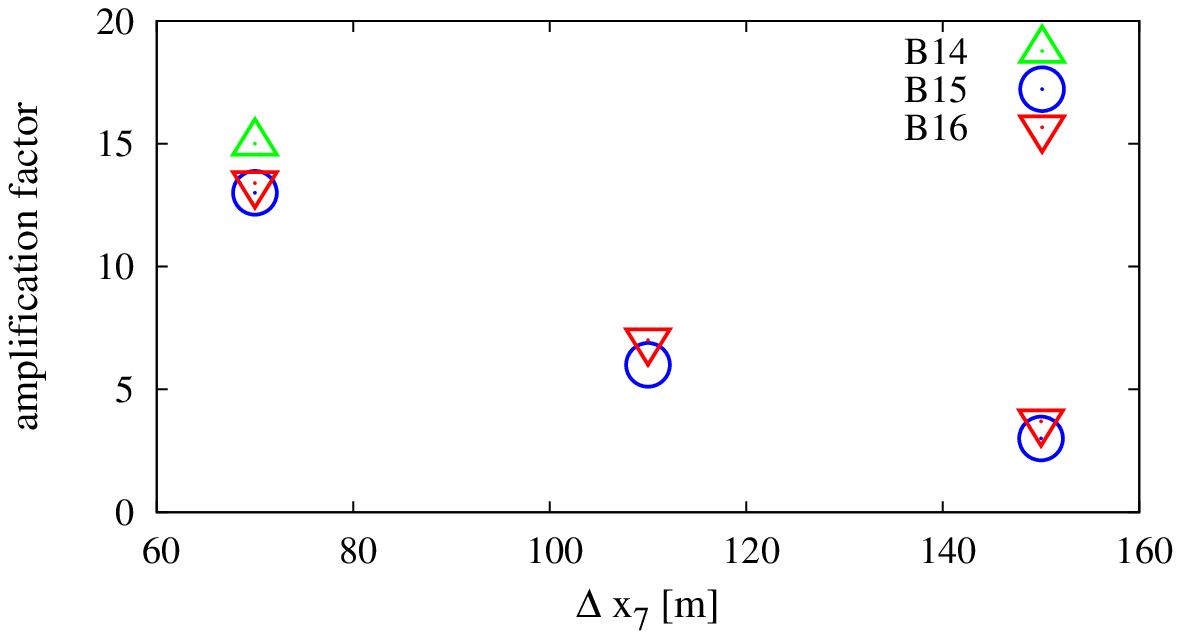}
\vspace{-10mm}
\end{center}
\caption{\label{fig2} (Top) The total magnetic-field energies as a function
  of time for H4B15 runs with three grid resolutions (B15-70m, B15-110m. B15-150m),
  for H4B14d70 (B14-70m), and for H4B16d70 (B16-70m). 
  The thin vertical lines denote the formation time of the
  BH. ${\rm E_B}$ is calculated by a volume integral only outside the BH horizon. 
  (Bottom) The dependence of the amplification factor of the
  maximum toroidal magnetic field in the merger on the grid resolution
  for all the models.}
\end{figure}

Figure~\ref{fig2} plots the magnetic-field energy as a function of time
for H4B15 runs, H4B14d70, and H4B16d70. Soon
after the onset of the merger, the magnetic-field energy is steeply
amplified because the KH vortices develop in the shear layer.  The
growth rate is higher for the
higher-resolution runs, because the growth rate of the KH instability is
proportional to the wave-number and hence the smaller-scale vortices
have the larger growth rate. We analyze the maximum magnetic-field
strength and plot the amplification
factor in the merger as a function
of $\Delta x_7$ in the lower panel of Fig.~\ref{fig2}. 
This clearly shows that the amplification factor
depends on the grid resolution but not on the initial magnetic-field
strength. This is consistent with the amplification mechanism due to the
KH vortices and qualitatively consistent with the local shearing-box 
simulation in Ref.~\cite{Obergaulinger:2010}. 
The magnetic-field energy at $t-t_{\rm mrg}\approx 5$\,ms in the
high-resolution run is 40--50 times as large as that of the
low-resolution run.

\begin{figure*}[t]
\begin{minipage}{0.35\hsize}
\begin{center}
\includegraphics[width=7.0cm,angle=0]{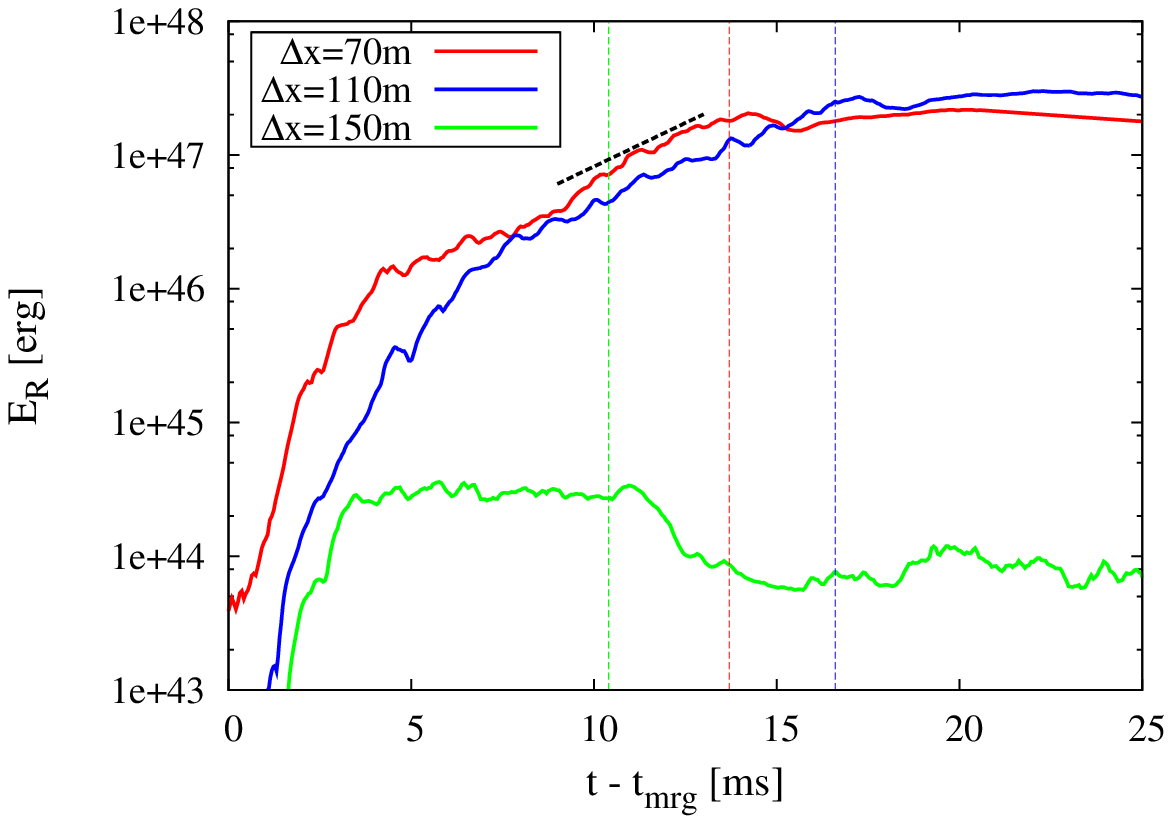}
\end{center}
\end{minipage}
\hspace{20mm}
\begin{minipage}{0.35\hsize}
\begin{center}
\includegraphics[width=7.0cm,angle=0]{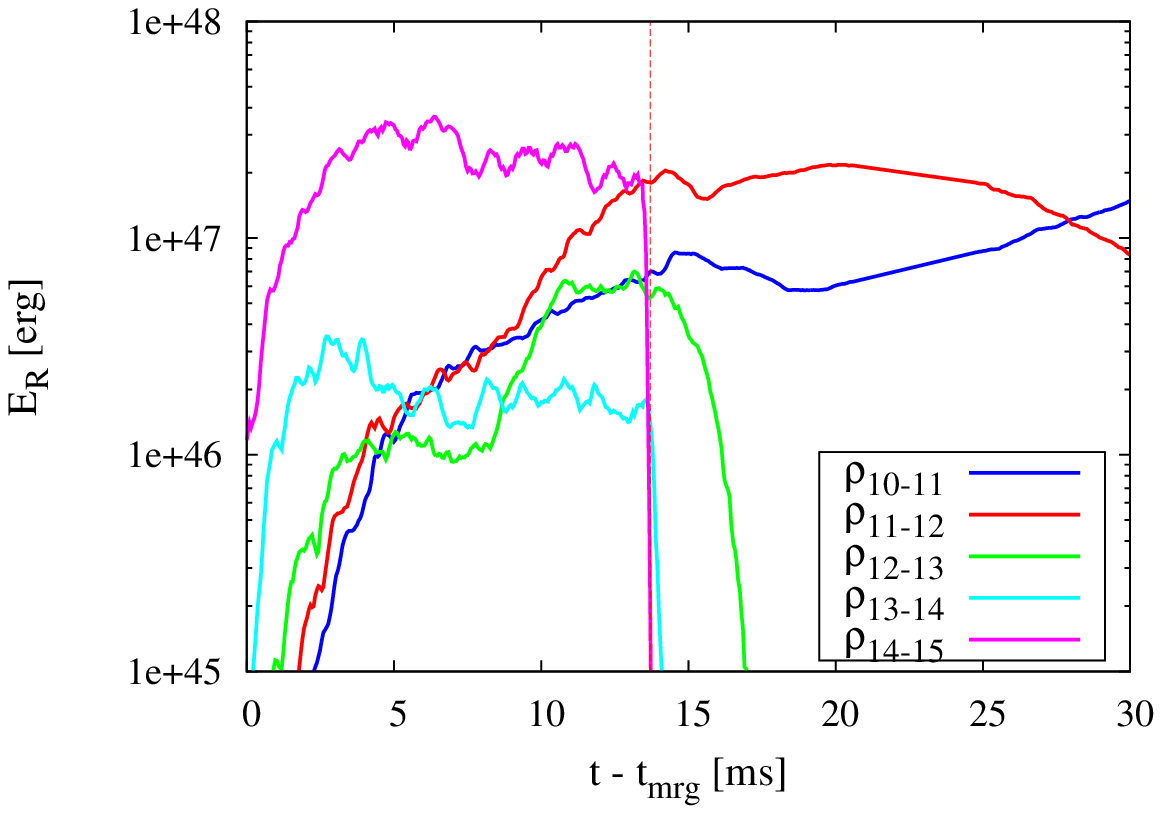}
\end{center}
\end{minipage}
\caption{\label{fig4} (Left) The magnetic-field energy of a radial
  component in the range $10^{11}{\rm g/cm^3}\le \rho \le 10^{12} {\rm
    g/cm^3}$ for H4B15 runs.  The thin vertical lines show the BH
  formation time.  The black-dashed line is an exponential function 
  $\propto {\rm e}^{2\sigma(t-t_{\rm mrg})}$ with $\sigma \approx 140$ Hz
  (see text in details).  
  (Right) The magnetic-field energy of the radial
  component 
  in $10^{a}{\rm g/cm^3}\le \rho \le 10^{a+1} {\rm g/cm^3}$
  for H4B15d70 with $a=10,11,12,13$ and 14. 
  The thin vertical line is the BH formation time. 
  } 
\end{figure*}

In the HMNS stage, the magnetic-field strength grows significantly in
the high- and middle-resolution runs but not in the low-resolution
run. We analyze the field amplification by foliating the HMNS in
terms of the rest-mass density, i.e., calculating the magnetic-field energy 
for $\rho_1 \le \rho \le \rho_2$ varying $\rho_1$ and
$\rho_2$. The left panel of Fig.~\ref{fig4} plots 
magnetic-field energy of a radial component 
for H4B15 runs with
$\rho_1= 10^{11}{\rm g/cm^3}$ and $\rho_2 = 10^{12}{\rm g/cm^3}$. 
We find that it grows in the middle- and high-resolution runs 
but not significantly in the low-resolution run.
We also find the high- and middle-resolution runs satisfy the criterion
$\lambda_{\rm MRI}^{\varphi}/\Delta x_7 \ge 10$ where $\lambda_{\rm
  MRI}^{\varphi}$ is the MRI wavelength of the fastest growing mode
for the ${\it toroidal}$ magnetic field, whereas the low-resolution run does not satisfy this criterion. 

We fit the growth rate of the magnetic-field energy by $\propto {\rm e}^{2\sigma (t-t_{\rm mrg})}$ 
for $ 8 \lesssim t - t_{\rm mrg} \lesssim 14$\,ms for the high-resolution run 
and find that  $\sigma \approx 140$\,Hz 
(for the middle-resolution run, it is $\approx 130$\,Hz for $8 \lesssim t - t_{\rm mrg} \lesssim 16$\,ms) 
which is several percents of the rotational frequency. 
This frequency agrees approximately with that of the nonaxisymmetric
MRI~\cite{Balbus:1992}. The right panel of Fig.~\ref{fig4} plots the 
magnetic-field energy 
in various density ranges for
H4B15d70.  This figure shows that the magnetic field in a high-density
region $\rho \agt 10^{13}\,{\rm g/cm^3}$ does not exhibit the
significant growth contrary to that in the lower-density region shown
in the left panel. This is because the MRI wavelength is proportional to 
$\rho^{-1/2}$, and hence, the wavelength in the high-density region ($\rho \agt
10^{13}\,{\rm g/cm^3}$) is too short to be resolved
even in our highest-resolution run. 
The growth rate in the range $10^{11} \lesssim \rho \lesssim 10^{12} {\rm g/cm^3}$ 
is greater than that in the range $10^{10} \lesssim \rho \lesssim 10^{11} {\rm g/cm^3}$ for $8 \lesssim t-t_{\rm mrg} \lesssim 14$ ms because the orbital angular velocity is larger in the higher density region. 
The same analysis shows that the magnetic fields are amplified even in $10^{13}{\rm g/cm^3}\lesssim \rho \lesssim 10^{14}{\rm g/cm^3}$ for H4B16d70 and not in $10^{12}{\rm g/cm^3}\lesssim \rho \lesssim 10^{13}{\rm g/cm^3}$ for H4B14d70. 
We conclude that the growth of
the magnetic-field energy in the HMNS phase is attributed to the ${\it
  nonaxisymmetric}$ MRI in the low-density region.  The magnetic
winding contributes to the growth of the toroidal magnetic-field
energy as well.

The MRI in the HMNS phase greatly 
amplify the magnetic fields. At the BH formation, the magnetic-field
strength is already saturated in
the high- and middle-resolution runs as found in Fig.~\ref{fig2}, and
thus, it does not much increase in the accretion torus formed after the
HMNS collapses.  On the other hand, the magnetic field is
still amplified in the accretion torus in the low-resolution run. This
is attributed to the insufficient resolution to capture 
the MRI in the merger and HMNS phases. 
Previous simulations often reported this picture due to the insufficient
resolution.  However, the picture we show in this paper is qualitatively
different from it.  The growth of the magnetic-field energy inside the
accretion torus is also seen for the low magnetic-field model H4B14d70
in Fig.~\ref{fig2} because the wavelength of the fastest growing mode of
the MRI in the HMNS is rather short in this model.  On the other hand,
the magnetic-field energy for H4B16d70 saturates at the formation of the
torus in Fig.~\ref{fig2}.  In reality, the magnetic-field energy may
reach the equipartition to the kinetic energy at the merger and inside
the HMNS.

\emph{Summary and discussion.}--- We have reported the results of
longterm and high-resolution MHD simulations of the BNS merger performed
in numerical relativity on the K.  The grid
resolution employed is highest among the simulations carried out so far.

We have found the KH vortices, which develop in the shear layer at the
onset of the merger, significantly amplify the magnetic-field strength
in a dynamical timescale. This feature can be found
only by a simulation with the grid spacing of $\lesssim$ 100m.

After the formation of a HMNS as a remnant of the merger, the MRI
amplifies the magnetic fields in the HMNS. Because the toroidal
magnetic fields are dominant in the HMNS, nonaxisymmetric MRI plays a
central role in amplifying the magnetic-field strength in this phase.

The HMNS eventually collapses to a BH surrounded by an accretion torus 
after the substantial angular-momentum transport inside it. 
Due to
the amplification mechanisms discussed above, the accretion torus
formed is strongly magnetized even at its formation.  The
magnetic-field energy is already saturated, and hence, does not
exhibit any remarkable growth. This indicates that 
a central engine of SGRBs would be modeled by a magnetized 
accretion torus with saturated strength.

Even after the longterm evolution, the global structure of the magnetic
field is toroidal-field dominant, and any coherent structure of the
poloidal component is not found. This does not agree with the previous
finding~\cite{Rezzolla:2011da}. Our results indicate
that the coherent poloidal field is not likely to be generated in
several 10\,ms after the BH formation, because the ram pressure
of the fall-back fluid elements toward
the BH and torus is quite strong and hence the outflow motion, which is
necessary to generate the poloidal component, is suppressed. This
implies that a new mechanism, which enhances the poloidal motion, is
necessary.

\emph{Acknowledgments.}---We thank K. Nitadori and T. Ishiyama for the optimization on K.  Numerical
computations were performed on the supercomputer K at AICS, XC30 at
CfCA of NAOJ, FX10 at Information Technology Center of Tokyo
University, and SR16000 at YITP of Kyoto University.  This work was
supported by Grant-in-Aid for Scientific Research
(24244028, 25103510,25105508,24740163,25103512,23740160), for Scientific Research on Innovative
Area (24103001), by HPCI Strategic Program of Japanese MEXT (hpci130025,140211). 
K. Kyutoku is supported by JSPS Postdoctoral Fellowship for Research Abroad.



\end{document}